\documentclass{natureclass}
\pdfoutput=1

\usepackage{lscape}
\usepackage{longtable}
\usepackage{amssymb}
\usepackage{amsmath}
\usepackage{url}
\usepackage{bm}
\usepackage{subcaption}
\usepackage{bbm}
\usepackage[ruled,vlined]{algorithm2e}
\usepackage{graphicx}
\usepackage{float}
\usepackage{icomma}
\def\BIISQ{{\sc biisq}}

\title{BIISQ: Bayesian nonparametric discovery of Isoforms and Individual Specific Quantification}

\author{Derek Aguiar$^{1}$, Li-Fang Cheng$^2$, Bianca Dumitrascu$^3$, Fantine Mordelet$^4$, \\Athma A Pai$^{5}$, \& Barbara E Engelhardt$^{1,6}$}

\begin{document}

\maketitle

\begin{affiliations}
 \item Department of Computer Science, Princeton University, Princeton, NJ
 \item Department of Electrical Engineering, Princeton University, Princeton, NJ
 \item Lewis Sigler Institute, Princeton University, Princeton, NJ
 \item Institute for Genome Sciences and Policy, Duke University, Durham, NC
 \item Department of Biology, Massachusetts Institute of Technology, Cambridge, MA
 \item Center for Statistics and Machine Learning, Princeton University, Princeton, NJ
\end{affiliations}

\begin{abstract}
Most human protein-coding genes can be transcribed into multiple possible distinct \emph{mRNA isoforms}. 
These alternative splicing patterns encourage molecular diversity and dysregulation of isoform expression plays an important role in disease etiology. 
However, isoforms are difficult to characterize from short-read RNA-seq data because they share identical subsequences and exist in tissue- and sample-specific frequencies. 
Here, we develop \BIISQ{}, a Bayesian nonparametric model to discover Isoforms and Individual Specific Quantification from RNA-seq data.
\BIISQ{} does not require known isoform reference sequences but instead estimates isoform composition directly with an isoform catalog shared across samples. 
We develop a stochastic variational inference approach for efficient and robust posterior inference and demonstrate superior precision and recall for short read RNA-seq simulations and simulated short read data from PacBio long read sequencing when compared to state-of-the-art isoform reconstruction methods.  
\BIISQ{} achieves the most significant gains for longer (in terms of exons) isoforms and isoforms that are lowly expressed (over 500\% more transcripts correctly inferred at low coverage in simulations).   
Finally, we estimate isoforms in the GEUVADIS RNA-seq data, identify genetic variants that regulate transcript ratios, and demonstrate variant enrichment in functional elements related to mRNA splicing regulation.
\end{abstract}

\section*{Introduction}

\begin{figure*}
 \begin{center}
 \includegraphics[width=10cm]{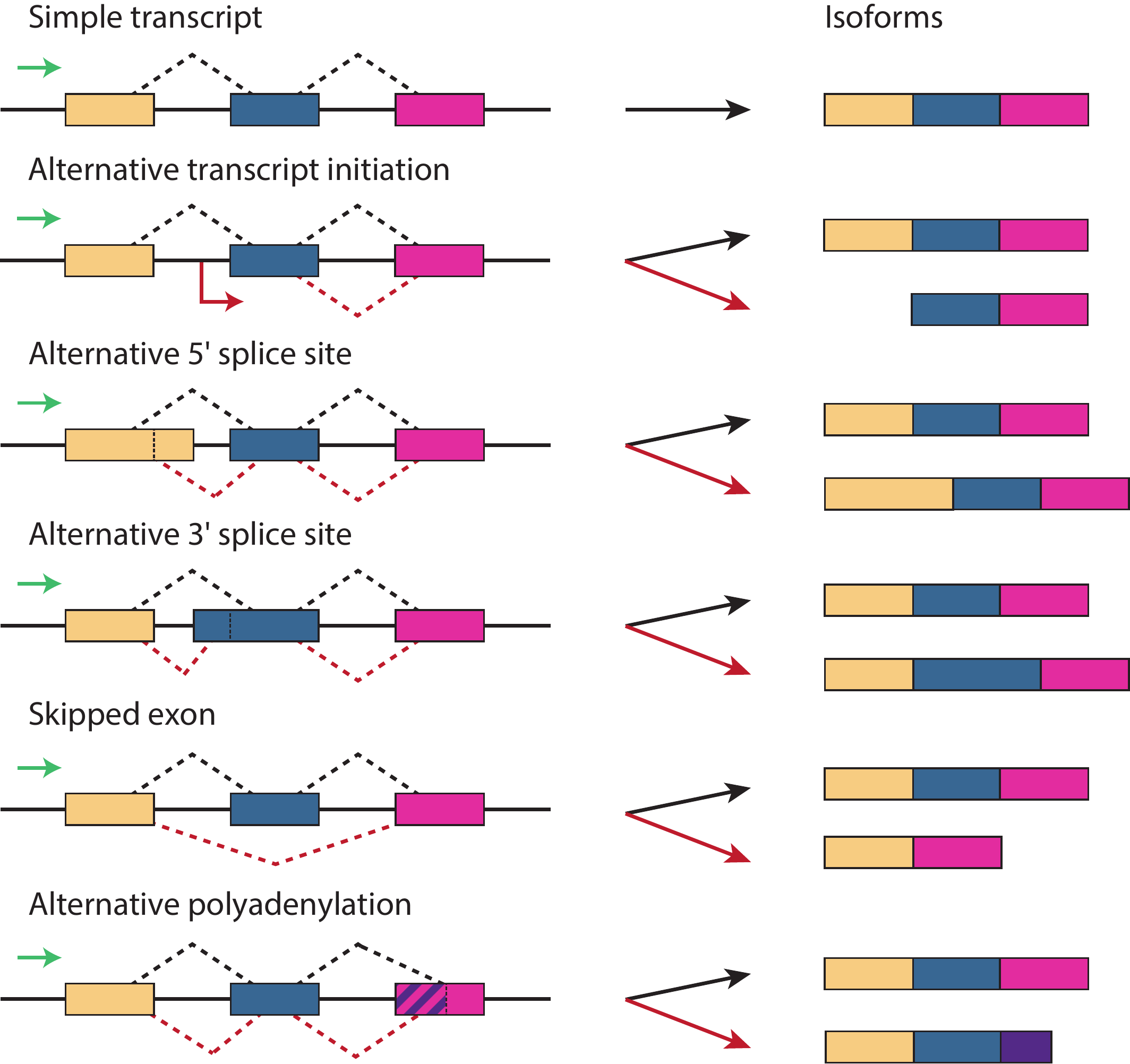}
\caption{{\bf A single gene may be transcribed into several distinct mRNA variants called \emph{isoforms} 
through alternative splicing mechanisms.} This figure shows six common types of splicing events (top to bottom): simple transcript; alternative transcription start site; alternative 5' splice site; alternative 3' splice site; skipped exon; and alternative polyadenylation.}
\label{fig:splicing}
\end{center}
\end{figure*}

Alternative splicing is the process by which a single protein-coding gene produces distinct mRNA transcripts, which vary in usage of component exons~\cite{Dutertre2010Alternative}.
Isoforms can differ by alternative transcription initiation sites, alternative usage of splice sites (either 5’ donor or 3’ acceptor sites), alternative polyadenylation sites, or variable inclusion of entire exons or introns (Figure~\ref{fig:splicing}). 
Altogether, alternative splicing enables the large diversity of mRNA expression levels or proteome composition observed in eukaryotic cells, which is particularly important for regulating the context-specific needs of the cell~\cite{Wang2008Alternative}.

It is estimated that 95\% of human protein-coding genes produce alternatively spliced mRNAs~\cite{Dutertre2010Alternative}. 
These splicing decisions are important drivers of many biological processes, with considerable variation in splicing patterns across human tissues~\cite{gtex2015}.
For example, mutations in splicing regulatory elements leading to disease pathogenesis and progression~\cite{Dutertre2010Alternative,Weber2008Molecular,Srebrow2006connection,tazi2009,faustino2003,li2016} and mutations in protein domains of specific splicing factors occur at a high rate in tumor cells, resulting in increased cellular proliferation~\cite{Venables2008Identification}. 
Furthermore, proteins resulting from splicing variants often have distinct molecular functions--for instance, the two variants of \emph{survivin} have opposite functions: one with pro-apoptotic and the other with anti-apoptotic properties~\cite{Vegran2007Association}.

Though there is increasing evidence of the biological importance of splicing processes, the precise role of alternative isoforms in regulating complex phenotypes is still largely uncharacterized. This gap in understanding is due, in part, to the difficulty of identifying and quantifying isoforms with high accuracy from RNA-seq data~\cite{mele2015}. 
Transcript reconstruction is essential to elucidate the role of gene expression in biological processes, because gene level quantification is convoluted by the multiple transcribed isoforms for each gene. The difficulties in isoform quantification stem from the tissue- and sample-specific composition and expression patterns of isoforms, the lack of a complete reference for isoform composition, and low abundance levels of many isoforms~\cite{Wang2008Alternative}. 
Further, RNA-seq reads that overlap informative splice junctions are rare, often noisy~\cite{Pickrell2010b}, and difficult to map to a reference genome~\cite{trapnell2009}. 
Improvements in reconstructing and quantifying tissue- and sample-specific isoforms would enable substantial improvements in understanding the role of alternative splicing in complex disease.

While many tools exist for isoform reconstruction using RNA-seq data, these methods have a number of drawbacks. 
First, many quantification methods assume that a high-resolution isoform sequence reference is available for each gene in the genome~\cite{Katz2010Analysis,griffith2010,beers}, which in practice is often not available or incomplete for non-model organisms and rare tissue or disease samples~\cite{mele2015}.
Second, most methods consider a single sample in isolation, which fails to exploit the sharing of isoforms across samples to gain power for identification of rare or low abundance isoforms~\cite{Trapnell2010Transcript,li2011,wei2011}. 
Third, many methods make technology dependent assumptions by controlling for specific biases (e.g., non-uniform sampling of reads~\cite{Li2010Modeling}) that do not generalize to mixtures of existing technologies or new technologies with different biases.
					
Our method, Bayesian isoform discovery and individual specific quantification (\BIISQ{}), addresses these limitations. 
First, \BIISQ{} uses annotations of transcribed regions as prior information~\cite{florea2005,harrow2012}, but the number and composition of isoforms across samples are estimated directly from the data, and the number of isoforms may grow with additional observations.
Second, \BIISQ{} explicitly captures isoforms shared across samples using a Bayesian hierarchical admixture model, which models multiple samples jointly and borrows statistical strength across samples to identify shared isoforms that may be in low abundance across samples.
Third, we assume that each nucleotide base in an isoform has an independent frequency in the mapped reads allowing \BIISQ{} to account for read mapping biases in RNA-seq data~\cite{Degner2009Effect}.

We develop a computationally tractable stochastic variational inference (SVI) algorithm to fit this model to RNA-seq data to estimate the structure of isoforms, probabilistically assign reads to isoforms, and compute sample-specific and global isoform proportions~\cite{Hoffman2013Stochastic}.
We compare and validate \BIISQ{} results on simulated data from the Benchmarker for Evaluating the Effectiveness of RNA-Seq (BEERS) software~\cite{beers} and from Pacific Biosciences (PacBio) Iso-seq data, which produces approximately $1-10$ Kb sequence reads potentially capturing full-length isoforms~\cite{pacbiodata}. 
Finally, we apply \BIISQ{} to a large RNA-seq data set from lymphoblastoid cell lines (LCLs)~\cite{Lappalainen2013} to identify the catalog of isoforms across samples. We use our catalog of sample-specific inferred isoforms and genotype data to identify genetic variants associated with isoform ratios and assess the functional significance of alternatively spliced genes and associated splicing variants.

\section*{Results}

The goal of isoform reconstruction and quantification is to robustly estimate both absolute and relative mRNA isoform expression levels, for both lowly and highly expressed transcripts, for each sample in a large RNA-seq data set from multiple sequenced samples. 
Our method, \BIISQ{}, approaches this problem by postulating a model of isoform composition and relative isoform expression shared across samples. 
Specifically, \BIISQ{} implements a Bayesian non-parametric hierarchical model of RNA-seq reads and isoforms inspired by the hierarchical Dirichlet process~\cite{teh2006hierarchical}, and we use stochastic variational inference (SVI) methods for computationally tractable and robust posterior inference~\cite{Hoffman2013Stochastic}.

The \BIISQ{} model probabilistically maps each RNA-seq read into a distribution over isoform exon compositions. 
Each sample is associated with its own distribution over isoforms, drawn from a global distribution over an arbitrarily large catalog of isoforms.
The exon composition of each isoform is modeled with a structured prior over exon usage that constrains the space of possible isoforms to those with support in the observed RNA-seq reads.
The model allows for extending the global isoform catalog by constructing novel isoforms given observed RNA-seq reads that could not have been generated from the current isoform catalog. 
Importantly, in our model, exon usage, RNA-seq read assignments, and the sample-specific and global isoform proportions are interpretable model parameters that translate directly to isoform composition and population- and sample-level isoform quantification (for details, see Methods). 

Variational methods enable computationally tractable posterior inference in Bayesian nonparametric models such as \BIISQ{} when applied to large genomic data~\cite{blei2016variational,gao2016}.
In brief, the posterior distribution of the \BIISQ{} model is intractable to compute directly; instead, we hypothesize a set of tractable variational distributions over the latent variables. Then, we iteratively compute the values of the variational distribution parameters that minimize the distance between the variational and true posterior distribution with respect to the Kullback-Leibler divergence~\cite{jordan1999,wainwright2008}. 
\BIISQ{} implements stochastic variational inference (SVI), an extension of variational inference that uses random subsets of the samples to compute each approximate update of the variational parameters~\cite{Hoffman2013Stochastic}. 
SVI was previously applied in eXpress to efficiently assign ambiguously mapped sequence reads for transcript abundance estimation~\cite{roberts2013}. 

\paragraph{Related isoform quantification methods.}

Methods for jointly inferring and quantifying alternatively spliced transcripts can be broadly partitioned into categories based on the required level of reference annotation~\cite{alamancos2014}. 
Transcriptome annotation dependent methods require complete annotation of the genome and transcriptome, including a description of isoform transcripts sequences and splice junctions~\cite{Katz2010Analysis,griffith2010,beers}.
In contrast, annotation-free methods require neither transcriptome nor genome annotations~\cite{Trapnell2010Transcript,grabherr2011,robertson2010}.
A third category of methods requires annotations of transcribed regions but is agnostic to isoform and splicing annotations~\cite{Trapnell2010Transcript,wei2011,li2011}; our method \BIISQ{} is in this category. 
Methods may exhibit characteristics shared across multiple of the three previously mentioned categories. 
For example, Cufflinks has evaluation modes that can be annotation-free or guided by reference annotations~\cite{Trapnell2010Transcript}.

We compared results from \BIISQ{} with three representative isoform reconstruction and quantification methods: Cufflinks~\cite{Trapnell2010Transcript}, CEM~\cite{wei2011}, and SLIDE~\cite{li2011}.
These methods were selected based on the following criteria: i) the ability to use annotations of transcribed genomic regions for isoform discovery and quantification, but no requirement for isoform transcripts or splice junction annotations; ii) coverage of combinatorial and statistical approaches; iii) support for both single-end and paired-end reads; and iv) high-quality performance in a recent benchmark study of isoform detection and quantification~\cite{angelini2014}.

Cufflinks uses a parsimonious approach to isoform discovery in order to find the minimal number of transcripts to explain the aligned reads~\cite{Trapnell2010Transcript}.
After filtering erroneous spliced read alignments, aligned reads are to assigned vertices in an \textit{overlap graph}, whose edges represent isoform compatibility between aligned reads.
Transcript assembly then reduces to finding a minimum set of paths through the overlap graph such that each aligned read is part of a path.
Transcript quantification uses a generative model for RNA-seq reads to compute a maximum \textit{a posteriori} estimate of the isoform quantifications, extending an earlier unpaired model~\cite{Hiller2009Identifiability}; confidence intervals are estimated using importance sampling.

CEM, an extension to the method IsoLasso~\cite{Li2011IsoLasso}, constructs a \textit{connectivity graph} to generate a set of candidate isoforms~\cite{wei2011}.
CEM and IsoLasso model the coverage of aligned reads at each location as a Poisson distribution and uses lasso regression to produce a set of inferred isoforms and abundance levels.
The principle difference between CEM and IsoLasso is the optimization procedure: CEM uses expectation maximization (EM) while IsoLasso solves a quadratic program. 
In our comparison, we preferred CEM because of the superior performance demonstrated on benchmark data~\cite{wei2012}.

The sparse linear modeling for isoform discovery and abundance estimation (SLIDE) method implements a statistical approach based on the start and end positions of aligned reads~\cite{li2011}.
SLIDE computes the number of aligned read start and end positions that group into transcribed regions of exons and organizes them into bins.
Isoform proportions are quantified using a linear model of the observed bin proportions; a modified lasso penalty limits the number and composition of isoforms.
In our methods comparison, we ran SLIDE using two distinct settings of the regularization parameter, $\lambda=0.01$ and $\lambda = 0.2$, which encourages more and fewer discovered isoforms, respectively.
We refer to results from SLIDE with these two parameter settings as SLIDE\_more and SLIDE\_fewer below.

\paragraph{Evaluation criteria.}

We evaluated precision and recall for each method in terms of exact and partial matches to simulated RNA-seq data~\cite{Steijger2013Assessment}.  
Precision and recall were calculated based on exact full length isoform matches between simulated and estimated isoforms (Equation \ref{eqn:pr}, Methods). 
\textit{Partial} precision and recall were calculated by defining imperfect matchings between each estimated transcript and the true transcripts (see Methods and Supplementary Fig. 1).
We controlled for issues regarding exon identification by counting an exon as successfully inferred if any subsequence of the inferred isoform overlapped an exon in the gene annotation.
Thus, reconstructing any subsequence of an exon was equivalent to reconstructing the whole exon correctly.

\paragraph{RNA-sequencing simulation: BEERS.}

The first evaluated our model on simulated data generated using the \emph{benchmarker for evaluating the effectiveness of RNA-Seq software} (BEERS)~\cite{beers}.
After removing genes with less than three exons, we divided RefSeq genes into three equally sized groups according to exon counts producing gene sets with $3-6$ exons, $7-12$ exons, and $13-312$ exons.
We then simulated $10,000$ single-end reads from $100$ samples and $35$ genes randomly selected from each group ($105$ simulated genes in total) using simulator parameters to vary read length and the number of distinct isoforms.

\begin{figure*}[h!]
 \begin{center}
\includegraphics[width=\textwidth]{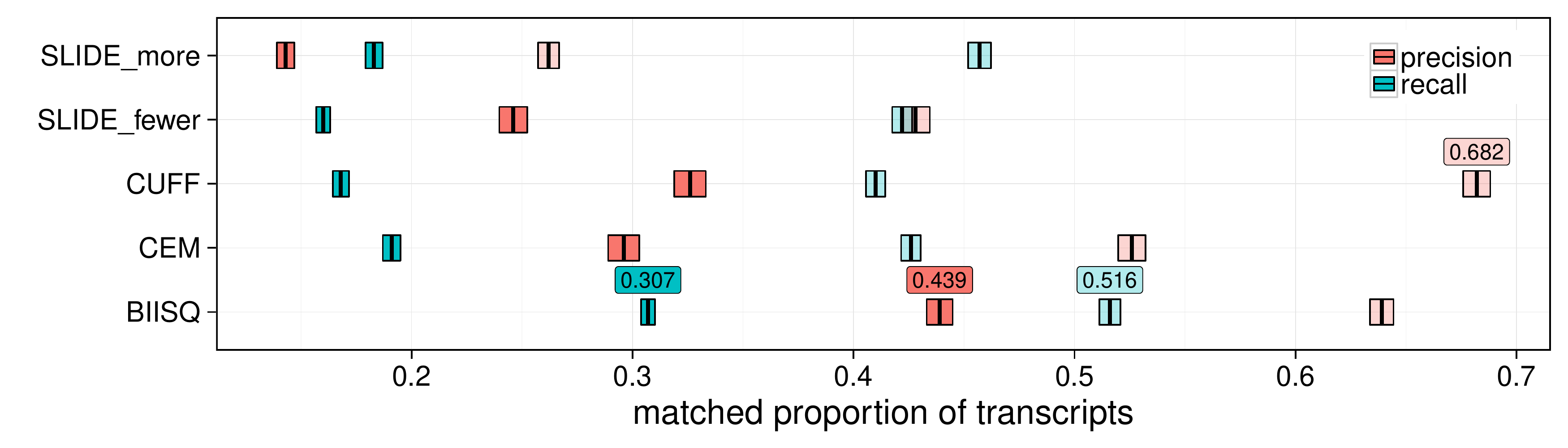}
\caption{\small {\bf Isoform discovery precision and recall for simulated data.} Precision (red) and recall (blue) of the results from \BIISQ{}, CEM, Cufflinks (CUFF), and SLIDE (SLIDE\_more and SLIDE\_fewer) applied to the BEERS simulated single-end RNA-seq data. 
The thick center bars denote the mean precision or recall and the fill denotes twice the standard error.
Transparent fill denotes partial precision and recall with a matching threshold of $0.2$. 
Across all methods, the best (partial) precision and recall values are annotated above their respective data points. 
} 
\label{fig:prerecexons}
\end{center}\vspace{-0.8cm}
\end{figure*}

To test the accuracy of each method, we applied the four isoform quantification methods to these simulated data and computed the precision and recall of the isoform discovery results---both perfect and partial matches (Figure~\ref{fig:prerecexons}). 
For perfect matchings, \BIISQ{} displayed significantly higher precision and recall across the $105$ genes (t-test, $p \leq 2.2\times 10^{-16}$). 
For partial matchings, \BIISQ{} showed significantly higher recall but lower precision than Cufflinks (t-test, $p \leq 2.2\times 10^{-16}$). 
The overall trend remains when considering the results factored by the number of exons in the gene or the number of alternatively spliced transcripts. However, there are opportunities to improve \BIISQ{} performance in precision and recall for highly spliced genes (Supplementary Fig. 2) and genes with a small number of exons (Supplementary Fig. 3).
These results suggest that \BIISQ{} makes marginal sacrifices in false discovery rate (FDR) to identify a higher proportion of true isoforms relative to Cufflinks.

Next, to investigate the source of \BIISQ{}'s increased precision and recall, we evaluated the number of perfectly inferred isoform transcripts for the most lowly expressed transcript in each gene. 
Across all samples in the simulated data, \BIISQ{}, CEM, Cufflinks, SLIDE\_more, and SLIDE\_fewer correctly inferred $3,011$, $1,205$, $996$, $1,167$, and $1,067$, respectively, out of a total of $10,500$ transcripts (i.e., $105$ genes and $100$ samples per gene) highlighting that much of the recall gains of \BIISQ{} originated from isoform transcripts with low expression levels.

\begin{figure*}
\centering
\includegraphics[width=\linewidth]{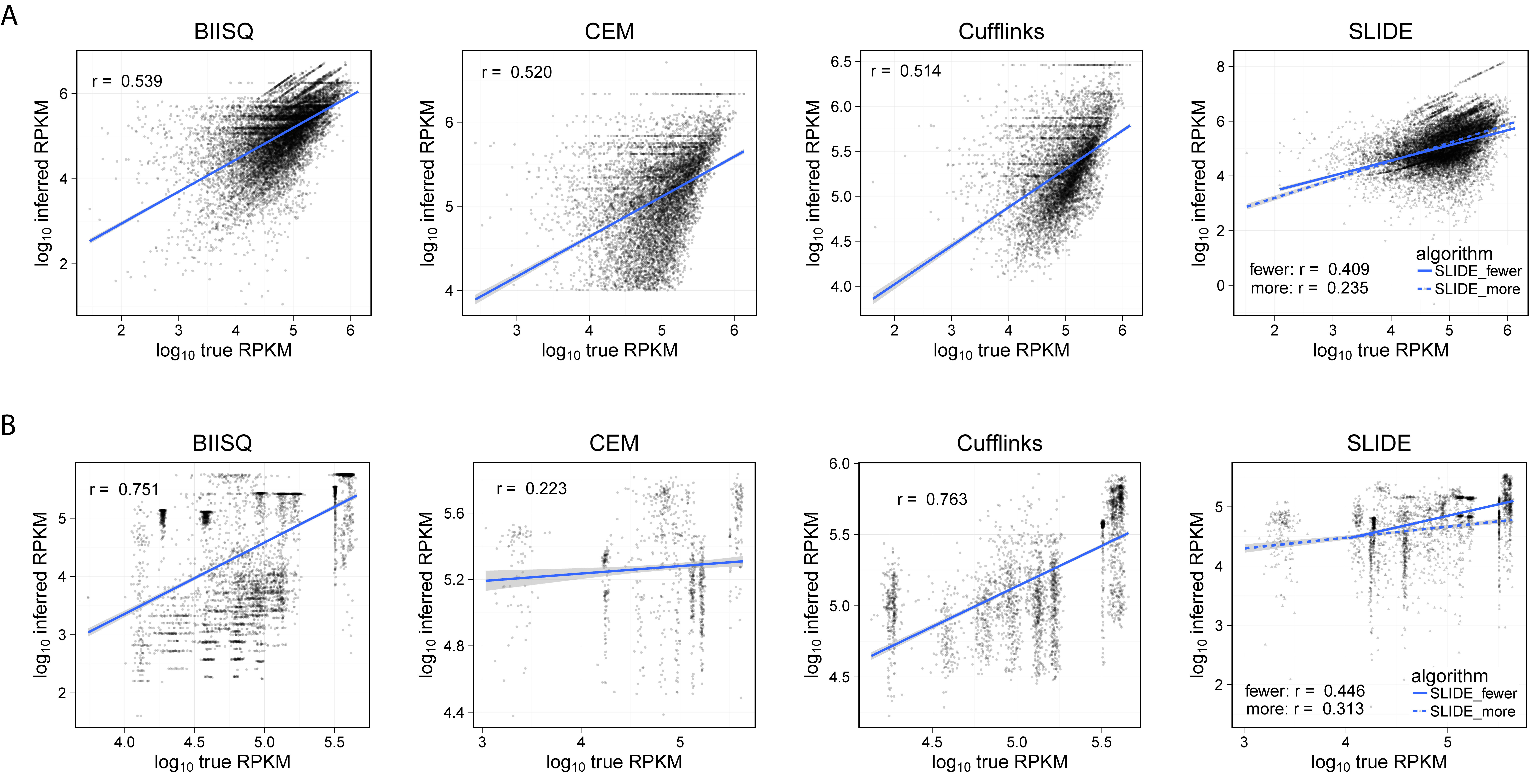}
\caption{{\bf Isoform quantification accuracy.} Correlation between true RPKM and inferred RPKM for (A) BEERS simulated data and (B) Iso-Seq simulated data. Pearson correlation coefficients for results from \BIISQ{}, CEM, Cufflinks, SLIDE\_more, and SLIDE\_fewer were (A) $0.539$, $0.520$, $0.514$, $0.235$, and $0.409$, respectively, for BEERS simulated data and (B) $0.751$, $0.223$, $0.763$, $0.313$, and $0.446$, respectively, for simulated data from Iso-Seq reads.}
\label{fig:quant}
\end{figure*}

We assessed the quantification accuracy of each method by computing the correlation between true and inferred normalized read counts (reads per kilobase of exon model per million mapped reads, or RPKM).
\BIISQ{} inferred positive expression for $13,607$ transcripts compared to $8,512$, $8,428$, $7,516$, and $7,062$ for SLIDE\_more, CEM, Cufflinks, and SLIDE\_fewer, respectively.
We found a wide range of expression level estimates from the four methods (Figure~\ref{fig:quant}A), which is typical of isoform quantification in human samples~\cite{Steijger2013Assessment}. 
Overall, \BIISQ{} showed the highest correlation of results across the BEERS data, followed by CEM and Cufflinks (\BIISQ{} $r=0.539$, CEM $r=0.52$, Cufflinks $r=0.514$).

We next investigated the characteristics of inferred transcripts with positive predicted expression by each method.
All methods inferred a similar number of transcripts when the number of exons was small~(at least three exons and at most six).
However, \BIISQ{} quantified over $2,300$ additional transcripts compared to Cufflinks, CEM, and SLIDE for medium length transcripts (total number of exons $\in [7,12]$) and at least $1,948$ additional transcripts for larger transcripts (total number of exons $\in [13,312]$; Supplementary Fig. 4). 
We also evaluated the number of transcripts quantified in terms of \emph{coverage}, or the number of bases sampled from the transcript with simulated reads normalized by the transcript length.
Consistent with our results on low frequency transcripts, \BIISQ{} correctly infers almost 500\% more transcripts across samples for isoforms with coverage $<5$ within each sample~(Supplementary Fig. 5). 
As coverage increases, the difference in the number of transcripts correctly inferred between \BIISQ{} and competing methods diminishes, with SLIDE\_more inferring the largest number of transcripts across all samples for transcripts at coverage $\geq 400$ within each sample. 
These results highlight the benefits of \BIISQ{}'s model based approach to isoform sharing across samples.

\paragraph{Short read RNA-seq simulations from long-read RNA-seq data.}

The BEERS simulated RNA-seq data models technology-specific biases of short read RNA-seq data but does not capture the exon composition of true transcript isoforms.
The Pacific Biosciences Iso-Seq protocol is a single molecule transcriptome sequencing technology that offers read lengths of up to $10$ Kb~\cite{pacbiodata}.
Iso-Seq reads may span entire RNA transcripts, making the characterization of isoform composition straightforward relative to inference from short-read data, where each read may map to one of many possible isoforms. 
Thus, long read sequencing provides a medium for experimentally driven evaluation of isoform reconstruction;
however, the cost and platform specific error rates make this technology infeasible to replace short read RNA-seq in the near future, necessitating the development of methods such as \BIISQ{} for short read RNA-seq data.
Further, while reconstruction is aided by long reads, quantifying isoforms is challenging due to the high costs at the relatively low throughput realized by Iso-Seq, making precision and recall the principle metrics for evaluation of Iso-Seq data~\cite{rhoads2015}.

We simulated short-read RNA-seq data from full length Iso-Seq reads, which allows us to precisely capture true isoform composition and proportions in simulated data. 
To do this, we constructed a reference set of genes from the Iso-Seq human transcriptome reference samples of heart and brain tissue.
After mapping genes and transcripts across tissues, we identified seven genes with two or more isoforms in the heart and brain tissues (see Supplementary Methods). 
Iso-Seq reads have a different error profile than other short-read technologies, so we evaluated both GMAP and STARlong's Iso-Seq read alignments to the human genome version hg19~\cite{pacbio_bestpractices} (Supplementary Figs. 6, 7). 
To sample paired-end short reads from the Iso-Seq reads, 
the simulator copied the RNA sequence from the Iso-Seq read and the alignment from GMAP or STARlong. 
For seven transcripts (Supplementary Table 1), we simulated reads with lengths $50$ bp, $100$ bp, and $200$ bp for $50$ replicates of brain and heart tissue Iso-Seq samples. 
We define transcript \emph{span} relative to the transcript sequence coverage of simulated reads; for example, a span of $0.5$ indicates that short reads were simulated until the simulated reads mapped to half the length of the Iso-Seq reads for a specific transcript. 

\begin{figure*}[h!]
 \begin{center}
\includegraphics[width=\linewidth]{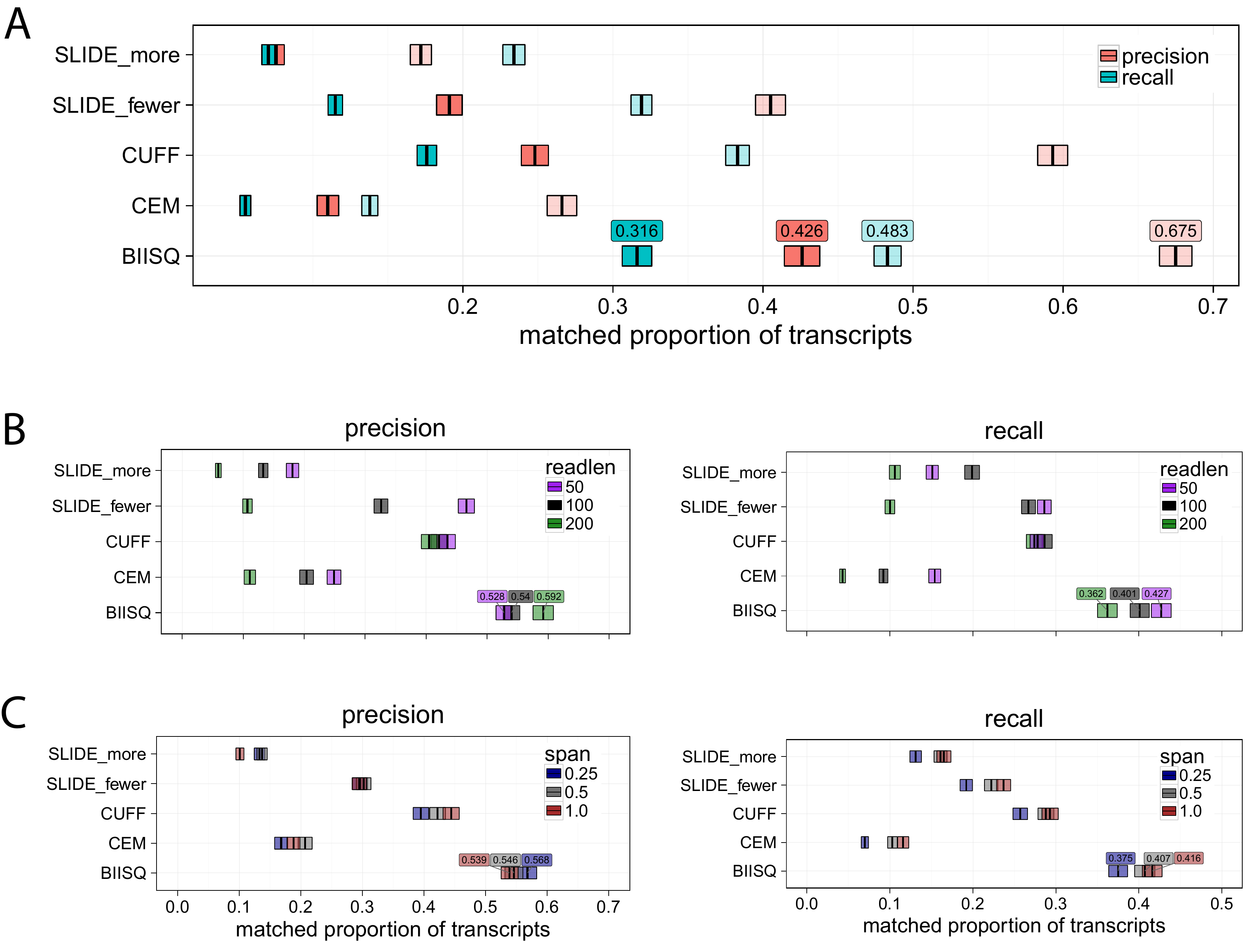} 
\caption{{\bf Comparison of methods on Iso-seq simulations.} Precision (red) and recall (blue) of the results from \BIISQ{}, CEM, Cufflinks (CUFF), SLIDE\_more, and SLIDE\_fewer applied to (A) the short-read data simulated from Iso-Seq reads; (B) simulated data split by read length; and (C) simulated data split by span. 
Transparent fill denotes partial precision and recall with a matching threshold of $0.2$
The thick center bars denote the mean precision or recall, and the fill denotes twice the standard error.
The best (partial) precision and recall values are annotated above their respective points.
}  \label{fig:pbres}
\end{center}\vspace{-0.8cm}
\end{figure*}

We first evaluated the accuracy of each method in terms of perfect and partial precision and recall.
\BIISQ{} achieved the highest precision and recall from both exact and partial matching thresholds across the seven Iso-Seq transcripts (Figure~\ref{fig:pbres}A).
This strong performance improvement remains when partitioning the RNA-seq data by read length and span (Figure~\ref{fig:pbres}B, C). 
Importantly, the performance of \BIISQ{}, Cufflinks, and SLIDE\_fewer does not deteriorate substantially in either precision or recall for paired-end short reads relative to the deterioration in performance from CEM and SLIDE\_more (Figures~\ref{fig:prerecexons} and~\ref{fig:pbres}A).

We then compared isoform quantifications across the four methods in the paired-end short read data (Figure~\ref{fig:quant}B).
\BIISQ{} computed positive quantifications for $5,595$ transcripts while Cufflinks, SLIDE\_fewer, SLIDE\_more, and CEM inferred $3,248$, $2,170$, $1,296$, and $1,025$ transcripts respectively.
Rankings of quantification results on paired-end data largely mirrored the BEERS simulations and, in total, \BIISQ{} and Cufflinks showed the highest average correlation across the BEERS and Iso-Seq data ($\bar{r}=0.645$, $\bar{r}=0.639$).
\BIISQ{} and Cufflinks also showed the highest agreement between any pair of distinct methods ($r=0.567$, Supplementary Fig. 8). 
We also found that BIISQ inferred more transcripts across all exon compositions~(Supplementary Fig. 9) and all spans~(Supplementary Fig. 10) in the Iso-Seq data.

In theory, paired-end reads are more informative for isoform reconstruction and quantification than single-end reads largely because reads sampled from the same fragment are transcript specific and may span additional non-contiguous junctions.
However, isoform transcripts often share many splice junctions, making them difficult to deconvolve in short-read paired end data, whereas dissimilar transcripts are easier to differentiate using unique splice junctions. 
The Iso-Seq simulated data included transcripts with a lower average normalized distance between isoforms ($0.471$) compared to the BEERS simulated data ($0.584$), making these gains in short-read paired-end data difficult to achieve.
This transcript similarity and complications in modeling the insert length of paired-end data may have contributed to the dramatic decrease in CEM's accuracy on the Iso-Seq simulated data~(Figure~\ref{fig:pbres}).  

We investigated the run time of each method as functions of the number of exons, gene length, read length, and span; \BIISQ{} run times are averaged across 20 runs and do not include the one-time preprocessing for converting aligned reads to read-terms (Supplementary Figs. 11-14).
CEM was the most efficient method tested, followed closely by Cufflinks, while \BIISQ{} and SLIDE had the longest run times.
However, isoform reconstruction can be parallelized at the level of reference transcript, so difficulties associated with running multiple iterations of \BIISQ{} on large data may be reduced by having many compute nodes to process distinct genes in parallel. 
Gene and read lengths had marginal effects on run time (Supplementary Figs. 11 and 12). 
However, CEM and Cufflinks showed increased run time as a function of the number of exons (p-value for linear regression of run time versus the number of exons for CEM and Cufflinks: $p \leq 8.79 \times 10^{-4}$ and $p \leq 1.64 \times 10^{-3}$, respectively) 
and the span (linear regression of run time versus span $p \leq 7.44 \times 10^{-5}$ and $p \leq 8.85 \times 10^{-5}$).
SLIDE\_all exhibited increased run time as a function of exons ($p \leq 6.16 \times 10^{-6}$) but no significant increase for span ($p \leq 0.32$).
In contrast, the run time of \BIISQ{} shows no dependence on the number of exons ($p \leq 0.61$) or span ($p \leq 0.29$) (Supplementary Figs. 13 and 14); this efficiency is likely due to modeling reads as read-terms and the aggregation of similar read-terms in the initial processing of the aligned sequence read input.

\paragraph{GEUVADIS RNA-seq data for $462$ samples.}

Next, we applied \BIISQ{} to RNA-seq data for $462$ lymphoblastoid cell lines (LCLs) from the GEUVADIS RNA sequencing project for 1000 Genomes samples~\cite{Lappalainen2013}. 
We built a model of transcription for each gene from the human genome annotations in GENCODE release 19 and mapped RNA-seq reads with STAR 2-pass to the human genome version hg19 (Methods).
Applying \BIISQ{} to these data, we discovered $31,712$ novel and $14,044$ known transcript isoforms with respect to the GENCODE database v19 transcript isoform annotations, considering only perfect matches to isoform exon composition.
When using a matching threshold of $0.2$, we discovered $24,871$ novel and $20,885$ known isoforms.
The distribution of the number of isoforms per gene is peaked for genes with no evidence of alternative splicing (one transcript) and heavily spliced genes ($\geq 7$ transcripts) although we note that this distribution could be confounded by erroneous splice junctions and fragmented transcripts in the \BIISQ{} output (Online Methods and Supplementary Fig. 15)~\cite{Pickrell2010b}. 

\begin{figure*}[h!]
 \begin{center}
\includegraphics[width=0.95\linewidth]{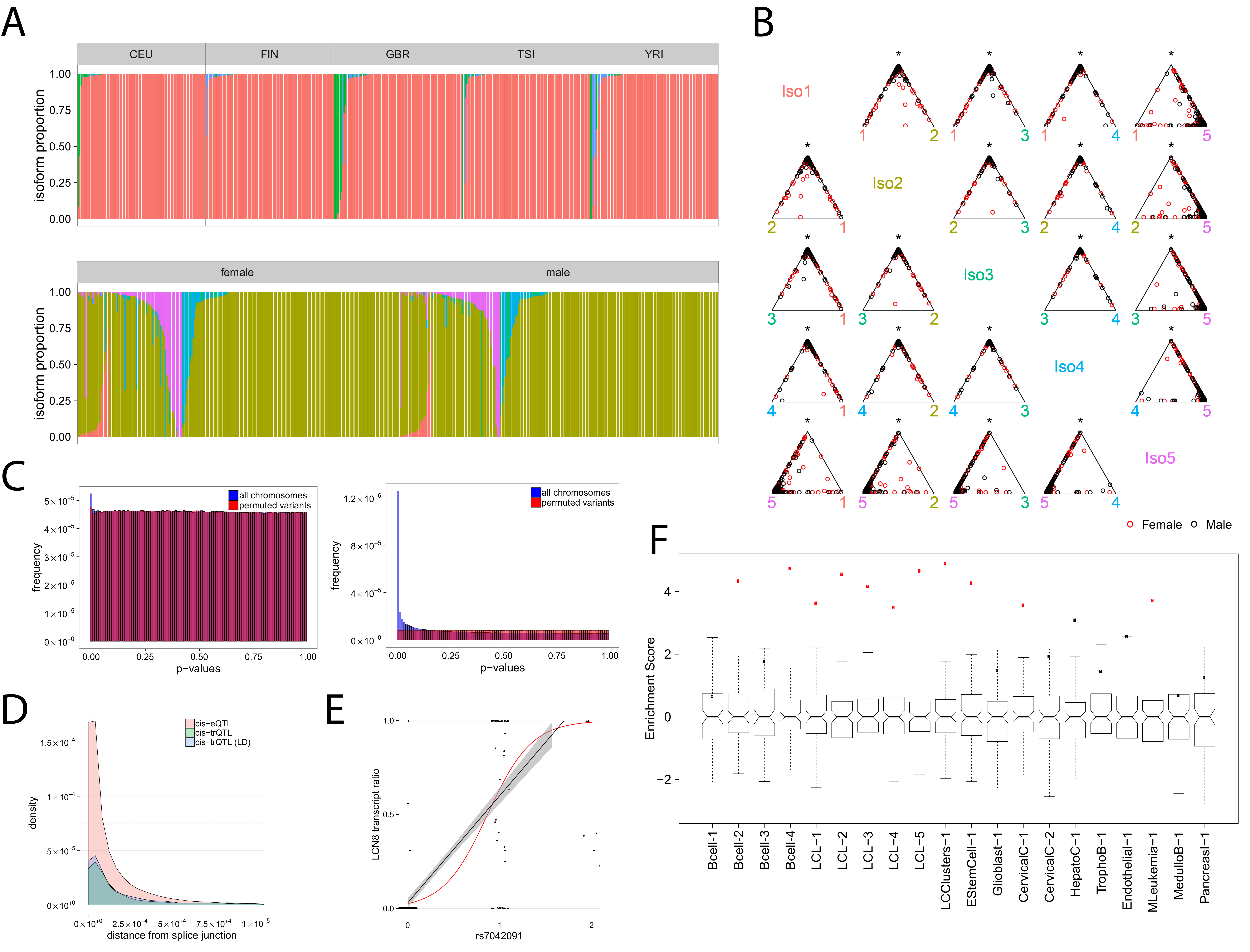}
\caption{{\bf Results for GEUVADIS data.} 
(A) The isoform quantification distribution where color denotes a unique isoform and each vertical bar is a single sample for genes \textit{PTPRN2} (top) and \textit{LGALS9B} (bottom).
(B) Simplex plots for gene \textit{LGALS9B} factored by sex. 
Each point (red for female, black for male) represents a sample's isoform composition distribution for the two isoforms denoted on the bottom axis and the remaining isoforms at the top intersection point.
(C) Matrix-eQTL p-value distribution for (left) cis-trQTLs and (right) cis-eQTLs. 
(D) The density of cis-trQTLs, LD pruned cis-trQTLs, and cis-eQTLs distances to the nearest canonical splice junctions in GENCODE. 
(E) \textit{LCN8} contained The most significant cis-trQTL ($p \leq 2.2\times 10^{-16}$). 
Linear and logistic regressions are shown in black and red. 
(F) Variant set enrichment scores (red points denoting significance). The x-axis includes cis-regulatory annotations (histone modifications and chromatin accessibility). 
} 
\label{matrixeqtlres}
\end{center}\vspace{-0.8cm} 
\end{figure*}

To investigate population and sex specific splicing patterns, we first analyzed transcript ratio quantification patterns across all genes in the GEUVADIS data. 
We considered global signatures of differential transcript ratio usage and did not find a significant difference in the average isoform transcript counts across sex ($\chi^2$ test, $p \leq 0.99$) or population ($\chi^2$ test, $p \leq 1$) when counts were aggregated across protein coding genes (Supplementary Fig. 15). 
We then computed population and sex specific transcript ratio distributions for each protein coding gene independently using likelihood ratio (LR) tests (see Online Methods). 
We found $924$ and $148$ genes that show population-specific and sex-specific transcript ratio distributions, respectively ($\chi^2$ test, Bonferroni-corrected $p \leq 0.05$; Supplementary Table 14).
Gene \textit{PTPRN2} showed the most significant differential effects of population on isoform ratios (LR test, Bonferonni-corrected $p \leq 2.2\times 10^{-16}$; Figure~\ref{matrixeqtlres}A, top).
Gene \textit{LGALS9B} showed the most significant differential effects of sex on isoform ratios (LR test, Bonferonni-corrected $p \leq 2.2\times 10^{-16}$; Figure~\ref{matrixeqtlres}A, bottom, B).
Most samples express at most two of the five isoform transcripts of \textit{LGALS9B}, but females show more highly variable isoform expression levels than males, in particular for isoform 5 (Figure~\ref{matrixeqtlres}B). 

Next, due to scarce prior information on population or sex specific transcript ratios, we validated these results by testing for over-representation of population (CEU, TSI, FIN, GBR, YRI), European (EUR) vs. African (AFR) (termed \textit{super population}), and sex specific variants in the exonic and intronic regions of the $924$ and $148$ genes.
We compared this set to variants in background genes defined as non-overlapping  gene regions in human genome version hg19.
To control for the correlation structure of variation throughout the genome, we preformed linkage disequilibrium (LD) pruning by removing variants in high $r^2$ with neighboring variation~(see Methods). 
We varied the minor allele frequency (MAF), population-, and sex-specific thresholds to test the hypergeometric test's sensitivity to the threshold. 
We found a significant over-representation of population-specific variants at MAF $\geq 0.15$  and a population threshold requiring $\geq 85\%$ of alleles in a sample to exist in a specific population (hypergeometric test, Bonferroni-corrected $p \leq 1.8 \times 10^{-5}$; Supplementary Table 2). 
We computed over-abundance tests for European or African specific variants (Supplementary Table 3) and find that they require a much lower threshold than the population tests (MAF=$0.05$, super population threshold=$0.75$, hypergeometric test, Bonferroni-corrected $p \leq 1.42 \times 10^{-5}$), which is expected given that most of the GEUVADIS sample is concentrated in the European population.
We found similar significance for sex specific variants (Supplementary Table 4) although assortment of variant alleles by sex follows more closely $Bernoulli(0.5)$ and thus the sex threshold is closer to $0.5$ (MAF=$0.05$, sex threshold=$0.60$, hypergeometric test, Bonferroni-corrected $p \leq 3.01 \times 10^{-4}$).

\clearpage

\paragraph{Quantitative trait loci analysis in GEUVADIS.}
Expression quantitative trait loci (eQTLs) are genomic regions that contain one or more genetic variants that influence mRNA expression~\cite{albert2015}; eQTL studies have been indispensable for identifying the functional significance of phenotype-associated variants from genome-wide association studies~\cite{nicolae2010}. 
While eQTL analyses capture the differences in gene expression, or an aggregate of all gene transcripts, relative to a genotype, more subtle relationships between genetic variation and each spliced transcript are obfuscated. 
We validated the RNA isoforms in GEUVADIS that \BIISQ{} identified by performing both cis-eQTL and QTL mapping of transcript ratios (trQTL) using high-resolution genotype data available for each of the GEUVADIS samples~\cite{Lappalainen2013,pickrell2010eqtls} (see Methods).

We found an enrichment of cis-trQTLs p-values compared to a uniform distribution of p-values obtained by permutations of the genotypes (Figure \ref{matrixeqtlres}C (left)).
We identified a total of $792$ cis-trQTLs (FDR $\leq 0.05$ after LD pruning) in $766$ unique genes. We found that the cis-trQTLs, and cis-eQTLs to a lesser degree, showed spacial clustering near splice junctions (Figure~\ref{matrixeqtlres}D). 
We found an enrichment of cis-eQTLs ($889,662$ in total) across $11,687$ unique genes (FDR $\leq 0.05$; Figure \ref{matrixeqtlres}C (right)).
A total of $511$ genes with a cis-trQTL also had a cis-eQTL ($66.7\%$) and $264$ cis-trQTLs were also cis-eQTLs ($33.33\%$).
These results suggest that cis-trQTL signals may be masked when restricting analysis to gene-level quantifications.

We then compared our results with the $419,983$ and $19,741$ cis-eQTLs in the EUR and YRI populations respectively, $116,079$ and $4,149$ cis-trQTLs, and $639$ unique cis-trQTL target genes (or \emph{eGenes}) identified in the GEUVADIS study~\cite{Lappalainen2013}.  
Only $17$ of the $766$ genes ($2.2\%$) that we identified as cis-trQTL eGenes were also annotated as eGenes in the GEUVADIS study. 
This result is likely caused by different processing protocols for the GEUVADIS data (e.g., the GEUVADIS study computed CEU and YRI QTLs separately) and the reference transcript methods used to quantify isoforms in the GEUVADIS data.
\BIISQ{} does not rely on known transcript isoforms and thus does not suffer from potential biases inherent in reference based isoform quantification.

We found significant correlation among \BIISQ{} cis-eQTLs and the GEUVADIS study cis-eQTLs (EUR: Supplementary Fig. 17; $\rho=0.76$ and $r_s=0.63$, t-test, $p \leq 2.2\times 10^{-16}$ and YRI: Supplementary Fig. 18; $\rho=0.56$ and $r_s=0.52$, t-test, $p \leq 2.2\times 10^{-16}$). 
There were not enough overlapping transcripts to compare \BIISQ{} cis-trQTLs and the GEUVADIS study cis-trQTLs, but, comparing the most significant cis-trQTLs for each gene between the \BIISQ{} and GEUVADIS study (EUR population) results did not show significant correlation (Supplementary Fig. 19; t-test, $p_\rho =0.49$ and $p_{r_s}=0.12$). 
Interestingly, \BIISQ{} cis-eQTLs were more highly correlated with \BIISQ{} cis-trQTLs (Supplementary Fig. 20; $\rho=0.74$ and $r_s=0.51$) than GEUVADIS study cis-trQTLs and cis-eQTLs (EUR: Supplementary Fig. 21; $\rho=0.3$ and $r_s=0.23$ and YRI: Supplementary Fig. 22; $\rho=0.18$ and $r_s=0.03$).

Next, we quantified potentially novel biological insight uniquely enabled by \BIISQ{}.
We computed genes with cis-trQTLs exclusively inferred by BIISQ compared to previous work~\cite{Lappalainen2013}.
The most significant cis-trQTL (rs7042091) was identified for gene \emph{LCN8} by \BIISQ{} (FDR $\leq 0.05$; Figure \ref{matrixeqtlres}E) both of which were not found in the GEUVADIS study (FDR $<0.05$)~\cite{Lappalainen2013}.
However, we found evidence for a cis-eQTL at this SNP-gene pair in the GTEx study across a number of tissues (most significant p-value in whole blood $p \leq 2.2 \times 10^{-16}$)~\cite{gtex2015}. 
Further, mutations in the \textit{SNCA} gene, typically expressed in neurons but also in LCLs, have been associated with Parkinson's disease and was among the $721$ uniquely inferred genes with a significant cis-trQTL~\cite{kobayashi2003}.
It has been demonstrated that alternatively spliced transcripts can cause protein misfolding~\cite{hegyi2011} and the misfolding of \textit{SNCA}'s protein product has been suggested as a therapeutic target for Parkinson's disease treatment~\cite{beraud2012}.
The synthesis of these results suggest the alternatively spliced transcripts of \textit{SNCA} might be interesting targets for future research and demonstrate the unique utility of \BIISQ{}.

To further characterize the functional relationships among these cis-trQTLs, we performed variant set enrichment (VSE) analysis for regions associated with variable intron splicing events identified by LeafCutter~\cite{li2016} and cis-regulatory elements (CREs) from ENCODE in a diverse set of cell types, including LCLs~\cite{hindorff2009,cowper2012,guo2016}. 
VSE is a statistical test that computes significant enrichment or depletion of an associated variant set (cis-trQTLs) co-localized among a set of genomic annotations. 
In particular, we computed enrichment of cis-trQTLs in sites associated with open chromatin; these data include: (a) enhancer and promoter-like regions identified through DNase I hypersensitive sites (DHS, including DNase I, H3K4me3, and H3K27ac) for B-cell samples; 
and a synthesis of DNase I hypersensitivity, formaldehyde-assisted isolation of regulatory elements (FAIRE), and chromatin immunoprecipitation (ChIP) experiments provided by the ENCODE project for (b) LCLs and (c) control cell types~\cite{encode2012} (see Supplementary Table 6 for details). 
We found that cis-trQTLs were significantly enriched in B-cells when considering only DHSs (Figure \ref{matrixeqtlres}F: \textit{Bcell-2,4}; VSE, Bonferroni-corrected $p \leq 2.07\times 10^{-4}$ and $p \leq 6.39\times 10^{-6}$) but not when jointly considering DHSs and H3K4me3 or DHSs and H3K27ac (Figure \ref{matrixeqtlres}F: \textit{Bcell-1,3}).
Furthermore, cis-trQTLs were significantly enriched for the combined DHSs, FAIRE, and ChIP LCL data (Figure \ref{matrixeqtlres}F: \textit{LCL-1}-5; VSE, Bonferroni-corrected $p \leq 3.60\times 10^{-3}$, $2.94\times 10^{-5}$, $2.19\times 10^{-4}$, $4.35\times 10^{-3}$, $2.16\times 10^{-5}$ respectively) and clusters of alternatively excised introns identified by LeafCutter (Figure \ref{matrixeqtlres}F LCClusters-1, VSE, Bonferroni-corrected $p\leq 1.57\times 10^{-5}$). 
We did not find significant enrichment in the control cell-types besides embryonic stem cell (VSE, Bonferroni-corrected $p \leq 2.36 \times 10^{-4}$), cervical carcinoma cells (VSE, Bonferroni-corrected $p \leq 3.38 \times 10^{-3}$), and mesoderm leukemia cells (VSE, Bonferroni-corrected $p \leq 2.87 \times 10^{-3}$) indicating that there may be significant sharing of cis-trQTL related chromatin markers between these samples and LCLs  (Figure \ref{matrixeqtlres}F: samples H1hescPk-1, Helas3Ifna4hPk-1, and K562Pk-1). 

Finally, we quantified enrichment of eGenes in the Database for Annotation, Visualization and Integrated Discovery (DAVID)~\cite{dennis2003} across nine annotation lists including the KEGG, SwissProt, UniProt, and InterPro databases (Supplementary Table 7). 
We compiled sets of high confidence isoforms for each gene by filtering out transcript isoforms not present in $\geq 10\%$ of the samples.
We computed functional enrichment for the remaining eGenes, and genes with $>1$, $>4$ and $>6$ transcript isoforms using all annotated human genes as the background set. 
We found cis-trQTL gene target enrichment in a single KEGG pathway, \emph{olfactory transduction} (BH adjusted $p \leq 2.5\times 10^{-3}$) (Supplementary Table 8). 
This pathway shows substantial transcript diversity: more than two thirds of olfactory receptors have been estimated to be alternatively spliced~\cite{Young2003}.
The most significant enrichment for the SwissProt and UniProt \textit{seq-feature} annotations across the $>1$, $>4$, and $>6$ gene sets were \textit{alternative splicing} and \textit{splice variant} respectively (BH adjusted $p \leq 2.2\times 10^{-16}$)  (Supplementary Tables 9-11). 
We found the most significant enrichment from InterPro to be protein kinases (BH adjusted $p \leq 2.2\times 10^{-16}$) which have been shown to exhibit high proteomic and functional diversity as the result of alternative splicing~\cite{Anamika2009}. 
These database enrichment results support \BIISQ{}'s ability to reconstruct biologically relevant alternatively spliced gene sets.

\section*{Discussion}

We presented a statistical model, \BIISQ{}, for quantifying RNA isoforms, which shares strength across samples to estimate isoforms (especially those at low abundance) without reference isoform compositions. 
We described a stochastic variational inference method for fitting \BIISQ{} to data that allows our approach to scale to genome-wide study data; further, \BIISQ{} showed increased efficiency as the coverage or size of the gene increased in paired-end Iso-Seq data. 
We demonstrated that our method improves substantially over three state-of-the-art methods in both precision and recall of isoforms on two different types of simulated data with significant improvement for low abundance transcripts. 
We applied \BIISQ{} to the GEUVADIS RNA-seq data and identified known and novel isoforms that we extensively validated, in part, by identifying cis-trQTLs that cluster near known splice junctions and are significantly enriched in cis regulatory elements associated with chromatin accessibility, histone modifications, and alternatively excised intron clusters.

\BIISQ{} has several advantages over existing representations of RNA isoforms: 1) sample-specific isoforms are drawn from a collection of global isoforms, which leads to higher power to discover low frequency isoforms by sharing strength across samples; 2) a Bayesian hierarchical approach enables the principled incorporation of high-quality prior information such as observed variation in the exon composition of isoforms; and 3) a nonparametric approach allows us to flexibly combine computationally tractable posterior inference with model selection (here, the number of isoforms). 
\BIISQ{} is guaranteed to converge to a local maximum, but the results on BEERS and Iso-Seq data demonstrate that the quality of isoform reconstruction is improved from taking the best (in the maximum a posteriori) solution from multiple random restarts.  

Our results on GEUVADIS data show that \BIISQ{} captures biologically interesting trends.
This is likely due to the fact that the potentially incomplete transcripts \BIISQ{} computes are still useful; in fact, many methods quantify alternatively spliced exons~\cite{Katz2010Analysis,anders2012} or excised introns~\cite{li2016} that correspond to incomplete transcripts.
This suggests that both the full-length and partial transcripts identified by \BIISQ{} can be biologically meaningful and considered for downstream analyses.
The flexible and robust model for isoform identification and quantification from short-read RNA-seq data in \BIISQ{} enables a more precise estimate of transcript isoform levels than is currently available, and opens the door to a better characterization of the cellular regulation and role of transcript isoforms in complex systems.

\section*{Online Methods}

\paragraph{The \BIISQ{} model.}

A gene is defined as an ordered list of contiguous transcribed exons or retained introns; we will refer to both DNA sequence types as \emph{exons} for simplicity.
A gene's exons are ordered from the \textit{5'} to \textit{3'} end of the gene, and gene transcripts are represented by an ordered list of integers denoting the exons included in that transcript. 
The model observations are $j = 1:m$ RNA-seq samples, each with $i = 1:n_j$ reads mapped to a reference genome or transcriptome (gene annotation).
The $i^{th}$ \textit{read term} includes the start and end of the mapped read sequence and the exons that are covered; paired reads are modeled as two connected read terms.
A gene includes $\iota = 1:E$ exons, and the set of global and sample-specific isoforms are indexed by $k = 1:K$ and $\ell = 1:L$ respectively. 
We represent the composition of an isoform as a binary vector where $1$ signifies an exon is included, and a $0$ encoded a spliced exon.

The \BIISQ{} model assumes read term counts $\{x_1, \dots, x_v,\dots, x_V\}$ that are generated from a multinomial distribution with probability vector determined by isoform $k$, which follows a Dirichlet distribution $\bm{\beta}_{k} \in \mathbb{R}^{V}$.
\begin{equation}
\bm{\beta}_{k} \sim Dirichlet(b_{k,1}\eta_1,\ldots,b_{k,V}\eta_V). 
\end{equation}
The function $b_{k,v}$ maps read terms to exons
\begin{equation}
\begin{aligned}
b_{k,v} & =  \begin{cases}
g(x_v,b'_{k,\iota})-\epsilon &\text{if } g(x_v,b'_{k,\iota})=1\\
g(x_v,b'_{k,\iota})+\epsilon &\text{if } g(x_v,b'_{k,\iota})=0
\end{cases} && \text{for } v \in \lbrace 1, 2,\cdots, V\rbrace \text{ and } \iota \in \lbrace 1, 2, \cdots, E \rbrace \\
\end{aligned}
\end{equation}
where $b'_{k,\iota} \sim \text{Bernoulli}(\pi_{k})$, for $\iota \in \lbrace 1, 2,\cdots, E\rbrace$, $\pi_{k} \sim \text{Beta}(r, s)$, and $g(x_v,b'_{k,\iota})=1$ if read term $x_v$ maps to the composition of isoform $k$ (Supplementary Methods). 
The hyperparameter $\epsilon$ adds uncertainty to the read term components of $\bm{\beta}_k$. 
$\bm{\beta}_{k}$ has a degenerate Dirichlet distribution, which is defined over the sub-simplex given by the mapping function $b_{k,v}$; this exon-to-read term mapping encourages sparsity over the read terms. 

The distribution of global isoforms follows a Dirichlet process with concentration parameter $\omega$ and a uniform base distribution $H = U_{\mathbb{N}}$ over the set of all isoforms, $G_0| \omega, H \sim \text{DP}(\omega, U_{\mathbb{N}})$. 
The set of natural numbers $\mathbb{N}$ defines the set of possible isoforms through their binary encodings; e.g., the number $5$ encodes the isoform with the first and third exons included in a three exon gene ($101$).
Sample-specific isoforms are distributed according to a Dirichlet process with base distribution $G_0$, and concentration parameter $\alpha$: $G_j| \alpha, G_0 \sim \text{DP}(\alpha, G_0)$.
The sharing of the base distribution $G_0$ ensures isoforms are shared among the samples and the clustering property of the Dirichlet process encourages observations to join existing isoforms (\textit{rich-get-richer} property). 
Sample-specific and global isoforms are related through a multinomial mapping variable $c_{j,l}$ and the latent isoform assignment for each read is drawn from a multinomial distribution $z_{j,i}$.
Finally, reads are drawn from a multinomial distribution with probability vector determined by the global isoform, $w_{j,i} \sim \text{Multinomial}(\bm{\beta}_{c_{j,l'}}), l'=z_{j,i}.$ See Supplementary Methods, Supplementary Fig. 24, and Supplementary Tables 12 and 13 for details of the \BIISQ{} model.

\paragraph{Posterior inference in \BIISQ{}.}
We developed a stochastic variational inference (SVI) method to tractably and robustly estimate posterior probabilities in the \BIISQ{} model, following prior work on SVI for the hierarchical Dirichlet process (HDP)~\cite{Hoffman2013Stochastic}. 
We modified this method for the \BIISQ{}-specific model parameters as follows: 
Variables $\textbf{$\pi$}_k$ and $\textbf{b'}$ were distributed as Beta-Bernoulli, which models the probability that an exon is included in isoform $k$.
The mapping function $g(\cdot)$ relates the exon inclusion variables to the global isoform Dirichlet distribution over read-terms. 
Sparsity in terms of the number of exons per isoform may be induced by controlling the hyperparameters of the Beta-Bernoulli hierarchy, which removes exons and reduces the probability of emitting the corresponding read-terms in the Dirichlet distribution through the mapping function $g(\cdot)$ (Supplementary Fig. 24).
Thresholds defined in the following description of the inference algorithm are configurable parameters.
 
To handle the expansion and contraction of the population-wide isoforms, we implemented a merge-propose-reduce step in SVI and executed this step every $30$ iterations~\cite{dahl2003,jain2004}. 
For every pair of isoforms, the \textit{merge} step calculates the likelihood of the data before and after merging the pair of isoforms; if the likelihood is greater after the merge for each sample, the merge is accepted. 
\BIISQ{} \textit{proposes} new isoforms by computing the union of exons in randomly sampled, poorly mapped read-terms, where the likelihood of that read mapping to existing isoforms is $<0.5$. 
If at least one novel isoform is proposed, \BIISQ{} reinitializes all variational parameters as defined by Supp. Algorithm 1.  
Finally, the \textit{reduce} step removes an isoform $k$ from the local and global distributions if, for all samples, there are no reads that map to isoform $k$ with a probability $>0.01$.

\paragraph{BEERS simulated data runs.} 
Single-end simulated RNA-seq reads were generated by the \emph{benchmarker for evaluating the effectiveness of RNA-Seq software} (BEERS)~\cite{beers}.
We divided RefSeq gene models into three equally sized groups according to exon count, producing groups of genes with 3-6, 7-12, and 13-312 exons.
For $35$ randomly selected genes in each group, we sampled $10,000$ error-free reads for seven parameter configurations: a fixed number of three novel transcripts 
with varying read lengths in $\{50,200,400\}$ and a fixed read length of $100$ with a varying number of novel transcripts in $\{2,3,4,6\}$ (see Supplementary Methods). 
For each gene model, we sample reads for $100$ samples drawn from the aligned RNA-seq reads in the BEERS simulated read pool. 
The start position of a read was sampled from a gamma distribution with a parameter that decreases linearly with the position to simulate a 5' bias.
The exonic composition of a novel splice form was generated by BEERS, and isoform proportions for each sample were sampled from a Dirichlet distribution with concentration parameter $1$.

\paragraph{Short read simulations from PacBio Iso-Seq long read data.}
We downloaded the full length non-chimeric human transcriptome liver, heart, and brain data from the Iso-Seq protocol, which included unaligned sequence reads and General Feature Format (GFF) reference files for each tissue~\cite{pacbiodata}.
The gene identifiers provided in the reference files were created independently for each tissue, so we constructed a reference set of genes and their transcripts across tissues as follows.
For each gene, we created a standard set of exons by parsing its transcripts and collapsing overlapping exons in the GFF files. 
We then mapped genes across the three tissues based on a base-pair overlap of 95\% and discarded non-unique mappings.
For each gene, we then mapped transcripts across tissues based on a 95\% overlap (see Supplementary Methods).
This process was conservative by design, leading to a confident baseline of cross-tissue isoforms. We found seven genes having at least two transcripts isoforms shared across at least two tissues (see Supplementary Table 1): \emph{BLOC1S6}, \emph{ZFAND6}, \emph{CYTH1}, \emph{APP}, \emph{C1orf43}, \emph{SPARCL1}, and \emph{RNF14}.
None of these genes were found to be expressed in liver and thus the liver data was discarded.

We mapped the Iso-Seq reads to the gene sequences of the identified transcripts from human genome version hg19 (Supplementary Table 1) using the GMAP and STARlong algorithms~\cite{pacbio_bestpractices}. 
We built an Iso-Seq short-read simulator (ISSRS) that simulates short-reads from longer Iso-Seq reads.
The inputs to ISSRS are sequencing parameters, a gene reference file with exon boundaries, and an aligned sequence read file. The outputs of ISSRS are aligned sequences in SAM format that contain shorter sequence reads but retain the read mapping biases present in the Iso-Seq data by copying the sequence position from the Iso-Seq reads.
In brief, the simulator works as follows: (1) compute Iso-Seq reads that map with high confidence to a known transcript; (2) for each read, determine the amount of sampling based on input coverage; (3) sample reads by attempting to add insert sizes distributed normally with mean 10 bp and 40 bp standard deviation; (4) output sampled reads while preserving sequence and quality scores from the aligned Iso-Seq transcripts in SAM and \BIISQ{} format (see Supplementary Methods).
For step (1), STARlong mappings yielded fewer false positives than GMAP, but GMAP produced many more usable alignments (Supplementary Figs. 6 and 7).
For the seven transcripts identified across tissues (Supplementary Table 1), we simulated reads with lengths 50 bp, 100 bp, and 200 bp and approximate coverage values of the input Iso-Seq transcripts of $0.25$, $0.5$, and $1$ for $50$ samples from brain and heart tissues.

\paragraph{GEUVADIS RNA-seq data preparation.}
RNA-seq reads from EBV-transformed LCLs were downloaded from the Genetic European Variation in Health and Disease (GEUVADIS) project~\cite{Lappalainen2013}.
\BIISQ{} requires read-terms -- mapped RNA-seq read start positions, end positions, and exons covered tuples -- and a model of transcription for each gene indicating contiguous transcribed subsequences including exons or retained introns.
To build the transcription model, we first extracted the protein coding representative (as defined by ENCODE annotation ``basic'') transcripts  
from the comprehensive gene annotations in GENCODE release 19 for human genome assembly version GRCh37.p13.
We then built a set of representative exons for each protein coding gene.
Most genes had a single transcript annotated as \textit{basic}; for the remaining genes, we kept the transcript with the largest number of exons.

To build the read-terms, we mapped the raw RNA-seq reads with STAR 2-pass to the human genome version 19.
We removed unmapped reads or non-primary reads that failed quality checks or were marked as duplicates. We computed the mapped reads that overlapped transcripts from our model, producing an intersection file.
The full catalog of read-terms is built from a first pass through the intersection files of each sample; we then construct read-term expression files for each sample from a second pass with the read-term catalog.
A final step reduces the number of read terms by collapsing terms with a similar start position and exon content to an approximate target number of read-terms of $2500$. 

\paragraph{Cis-QTL mapping.}
We used Matrix eQTL~\cite{shabalin2012matrix} to compute linear regressions and perform association mapping for local eQTLs and transcript ratio QTLs (cis-trQTLs) where the ratio of expression levels for each isoform to all isoforms in a gene---or the \emph{transcript ratio}---replaces the RPKM values for each gene~\cite{battle2014}. 
The logistic regression in Figure \ref{matrixeqtlres}E was computed using a generalized linear quasibinomial model.
We define the cis region of a gene as the genetic variants falling within 100 Kb of a gene's transcription start or end site.
Sex, population, the first three genotype principal components, and $15$ PEER factors estimated from the isoform ratio matrix were included as covariates using a standard processing pipeline for RNA-seq data to control for population structure and latent confounders~\cite{McDowell2016} (Supplementary Fig. 16 and Supplementary Table 5). 
The expression of cis-eQTLs were also quantile normalized and we removed genes with a single transcript or fewer than three exons in the computation of cis-trQTLs.
We generated the null distribution of p-values by permuting genotype labels while keeping isoform ratio labels constant~(Supplementary Methods). 
To achieve well calibrated null hypothesis p-values and filter transcripts containing false splice junctions, transcripts with ratios of 0 or 1 in all samples, as well as, transcripts expressed in less than 10\% of the samples were discarded.

\paragraph{GEUVADIS functional assessment.}

The Database for Annotation, Visualization and Integrated Discovery (DAVID v6.8, May 2016) analysis included functional enrichment for nine databases (Supplementary Table 7) and used the default whole genome set of genes~\cite{dennis2003}.
To reduce the affects of linkage disequilibrium (LD) on the variant set enrichment analysis, we computed LD blocks for YRI, CEU, FIN, GBR, and TSI populations with a minimum MAF of $0.001$, $r^2$ of $0.8$, and genotyping rate of $0.8$ using the rAggr interface to Haploview on the 1000 Genomes Project phase 3 data~\cite{Barrett2005}.
Variant set enrichment analysis was run on the LD blocks for ENCODE Encyclopedia 3 annotations: DNaseI hypersensitive sites, H3K4me3, H3K27ac, annotations generated from a chromatin state segmentation computational tool sourcing from the Broad Histone UCSC track for nine factors and nine cell types, DNaseI/FAIRE/ChIP synthesis annotations from ENCODE and OpenChrom~\cite{encode2012}, and LeafCutter clusters (Supplementary Table 6).
Sample Gm12878HMM was removed from enrichment analysis due to a non-normal null distribution (KS test, $p \leq 0.01$), which is required by VSE (Supplementary Fig. 23). 
We included annotations from LCLs as well as several other cell types as controls: glioblastoma, cervical carcinoma, hepatocellular carcinoma, trophoblast, and embryonic stem cells.

\paragraph{Population and sex specific splicing.}
Population and sex-specific transcript ratios were evaluated based on a likelihood ratio test. 
The alternative hypothesis modeled sample transcript ratios as draws from population and sex-specific Dirichlet distributions while the null hypothesis assumed a global Dirichlet distribution.
We computed the maximum likelihood estimates for parameters of the global, population-specific, and sex-specific Dirichlet distributions. 
Genes were selected for the isoform proportion plots (Figure \ref{matrixeqtlres}A) based on the likelihood ratio test 
\begin{equation*}
2log \left( \frac{\mathcal{L}(\hat{\theta}_{CEU}|x_{tr}^{CEU}),\mathcal{L}(\hat{\theta}_{FIN}|x_{tr}^{FIN}),\mathcal{L}(\hat{\theta}_{GBR}|x_{tr}^{GBR}),\mathcal{L}(\hat{\theta}_{TSI}|x_{tr}^{TSI}),\mathcal{L}(\hat{\theta}_{YRI}|x_{tr}^{YRI})}{\mathcal{L}(\hat{\theta}_{ALL}|x_{tr}^{ALL})} \right) \sim \chi^2
\end{equation*}
where $\hat{\theta}_{a}$ are the maximum likelihood estimates for the parameters of the Dirichlet distribution for population $a$, $x_{tr}$ are the sample transcript ratios and populations are denoted as superscripts.
Sex specific likelihood ratio tests were calculated analogously.

The EUR vs. AFR, population, and sex specific variant enrichment analyses were computed from the 1000 Genomes Project phase 3 main release data (human genome version hg19).
We describe the processing for population specific variants (sex specific and super population variants follow analogously). 
First, to control for linkage disequilibrium (LD), we masked variants in pairwise LD $>0.90$ using PLINK v1.9 (indep-pairwise 100000 1000 0.9).
We then extracted two orthogonal sets of variants in protein coding gene regions: population specific genes (selected set) and non-population specific genes (background set).
For a specific variant, let the count of an allele $a$ in population $p$ be denoted $|a|^p$, the set of all alleles be $A$ and the set of all populations $P$.
Then, a variant is population specific if
\begin{equation}
\exists a,p \mid \frac{|a|^p}{\sum_{q \in P} |a|^q} > t
\end{equation}
for some population threshold $t$.
Then, for each variant  we count the number of population specific alleles greater than a minor allele frequency threshold for our selected set and background set and test for an abundance of selected population specific alleles with a hypergeometric test.

\paragraph{Parameter settings for each method.}
To run \BIISQ{}, we initialized $K=1$, allowing the global distribution of isoforms for a given gene to consist of a single isoform expressing all exons.
For the BEERS simulations, we set \BIISQ{} hyperparameters to maximize precision and recall on a random held out RefSeq gene by grid-search.
The hyperparameters were as follows: $N-iter=5000$, $threads=1$, $use-cython=1$, $max-n-prop=3$, $min-n-prop=3$, $iter-prop=30$, $new-iso-prop=2$, $red-iso-prop=0$, $r=0.7$, $s=1$, $converge=1e-3$, $\alpha \in \{15,100\}$, $\omega \in \{8,50\}$, and $\eta \in \{8,50\}$. 
Solutions with the maximum likelihood were considered for evaluation. 
Hyperparameters for PacBio runs were the same as BEERS except $r=1.1$.
For GEUVADIS, hyperparameters were set to the same as PacBio with exceptions: $alpha=15$, $omega=8$, $eta=8$.

We ran related methods as follows:
\begin{itemize}
\item \textbf{GMAP v.2015-12-31 }~\cite{wu2005}:  \texttt{-D [data dir] -d pacbio -f samse -n 0 -t 16 --nofails [input fasta] > [output SAM] 2> [log file]}
\item \textbf{starLONG v.020201 }~\cite{dobin2012}: \texttt{--genomeDir [genome reference]} \texttt{--runThreadN 1} \texttt{--readFilesIn [input fasta]} \texttt{--outFileNamePrefix [output]}\\ 
\texttt{--runMode alignReads} \texttt{--outSAMattributes NH HI NM MD} \\
\texttt{--readNameSeparator space} \texttt{--outFilterMultimapScoreRange 1} \\
\texttt{--outFilterMismatchNmax 2000} \texttt{--scoreGapNoncan -20} \\
\texttt{--scoreGapGCAG -4} \texttt{--scoreGapATAC -8} \texttt{--scoreDelOpen -1} \\
\texttt{--scoreDelBase -1} \texttt{--scoreInsOpen -1} \texttt{--scoreInsBase -1} \\ 
\texttt{--alignEndsType Local} \texttt{--seedSearchStartLmax 50} \\
\texttt{--seedPerReadNmax 100000} \texttt{--seedPerWindowNmax 1000} \\
\texttt{--alignTranscriptsPerReadNmax 100000} \\
\texttt{--alignTranscriptsPerWindowNmax 10000}
\item \textbf{CEM v.2.6 }~\cite{wei2011}: \texttt{python runcem.py -x [BED reference] [input BAM] }
\item \textbf{Cufflinks v.2.2.1 }~\cite{Trapnell2010Transcript}:  \texttt{cufflinks  --library-type ff-firststrand -g  [GTF reference] -o [output] [input BAM]}
\item \textbf{SLIDE v.2012-02-17 (modification date)}~\cite{li2011}: \texttt{python slide.py [GTF reference] [BAM input] [GTF output] --read\_type mixed}
\end{itemize}

\paragraph{Evaluation criteria.}
An isoform transcript is defined by the set of exons that are expressed from a known gene reference. 
An evaluation criterion that requires the true and inferred exon sets to be identical is often conservative due to variable read coverage of exons. 
Therefore, isoform reconstruction was evaluated by considering both perfect and imperfect matchings to determine precision and recall (Supplementary Fig. 1).  
For exact matches, precision and recall were calculated based on exact full length isoform matches between true (simulated) and estimated isoforms: let true positives, false positives, and false negatives be denoted TP, FP, and FN respectively. 
Then,
\begin{eqnarray}
\label{eqn:pr}
precision=\frac{TP}{TP+FP} & & recall=\frac{TP}{TP+FN}
\end{eqnarray}
For inexact matches, \textit{partial} precision and recall were calculated by defining a matching $M$, or a set of pairs of inferred-true isoforms, that is of maximum cardinality and minimum weight (i.e., distance between isoform composition of a pair) between each computed transcript and the true transcripts as follows.
Let $K_C$ and $K_T$ be the set of estimated and true isoforms, respectively, which are Boolean vectors of length $E$ exons $\{1,2,...,E\}$. A $1$ at position $\iota$ signifies that exon $\iota$ is contained within that isoform, and $k[\iota]$ indexes the position of the Boolean vector $k$.
We define the distance between an inferred and true isoform $d_{k,l}$ for all $k \in K_C$ and $l \in K_T$ to be the Hamming distance.

The Hamming distance counts the number of mismatched exons between the estimated and true isoforms. 
The maximum cardinality minimum weight $M$ is then the solution to the optimization problem 
\begin{align}
\min M =& \sum_{k \in 1:K_C} \sum_{\ell \in 1:K_T} x_{k\ell} d_{k\ell} &\\
\textbf{s.t.} & \sum_{k \in 1:K_C} x_{k\ell}=1 & \forall \ell \in 1:K_T \label{truecond}\\ 
& \sum_{\ell \in 1:K_T} x_{k\ell}=1 & \forall k \in 1:K_C \label{inferredcond} \\
& x_{k\ell} \in \{0,1\} & \forall k \in 1:K_C, \forall \ell \in 1:K_T  
\end{align}
If the total number of isoforms is $I$, finding a maximum cardinality minimum weight matching can be solved in $O(I^3)$ time~\cite{edmonds1972}. %
If $d_{k\ell}$ is the distance between inferred isoform $k\in K_C$ and true isoform $\ell \in K_T$ for matching $M$, 
then $d_{k,\ell}=0$ ($d_{k,\ell}>0$) implies $k$ is a true (false) positive; if $d_{k,\ell}\leq p|E|$ ($d_{k,\ell} > p|E|$) then $k$ is a $p$-partial true (false) positive ($p$-TP and $p$-FP).  
Any true isoform not matched by a $p$-partial true positive is a $p$-partial false negative ($p$-FN). 
Using these definitions of $p$-TP, $p$-FP, and $p$-FN, we can compute $p$-precision and $p$-recall as in Equation~\ref{eqn:pr}.

\paragraph{Software.} The source code and software implementing the \BIISQ{} model and inference methods 
can be downloaded from: \url{https://github.com/bee-hive/BIISQ}.


\begin{addendum}
 \item DA was funded by NIH R01 MH101822. BEE was funded by NIH R00 HG006265, NIH R01 MH101822, NIH U01 HG007900, and a Sloan Faculty Fellowship. The authors gratefully acknowledge the GEUVADIS study and 1000 Genomes data online.
 \item[Author Contributions] DA, LC, BD, FM, AAP, and BEE contributed to developing the isoform reconstruction and quantification model. 
DA, LC, and FM designed and developed the code.
DA, LC, BD, and FM designed and implemented the computational analysis pipeline.
DA and BEE designed and performed the computational analyses.
DA, AAP, and BEE wrote the manuscript and all authors edited and approved the final manuscript. 
 \item[Competing Interests] The authors declare that they have no
competing financial interests.
 \item[Correspondence] Correspondence and requests for materials
should be addressed to DA~(email: daguiar@princeton.edu) and BEE~(email: bee@princeton.edu).
\end{addendum}

\vspace{10pt}

\bibliography{references}

\end{document}